Cite as:

### IEEE
M. Jafarzadeh, D. Hussey and Y. Tadesse, "Deep learning approach to control of prosthetic hands with electromyography signals", in *2019 IEEE International Symposium on Measurement and Control in Robotics (ISMCR)*, Houston, Texas, USA, 2019, pp. A1-4-1-A1-4-11.

### ACM
Jafarzadeh, M., Hussey, D. and Tadesse, Y., 2019. Deep learning approach to control of prosthetic hands with electromyography signals. In *2019 IEEE International Symposium on Measurement and Control in Robotics (ISMCR)*. IEEE, pp. A1-4-1-A1-4-11.

### MLA
Jafarzadeh, Mohsen et al. "Deep Learning Approach To Control Of Prosthetic Hands With Electromyography Signals". IEEE, *2019 IEEE International Symposium On Measurement And Control In Robotics (ISMCR)*. IEEE, 2019, pp. A1-4-1-A1-4-11.

### APA
Jafarzadeh, M., Hussey, D., & Tadesse, Y. (2019). Deep learning approach to control of prosthetic hands with electromyography signals. In *2019 IEEE International Symposium on Measurement and Control in Robotics (ISMCR)* (pp. A1-4-1-A1-4-11). Houston, Texas, USA: IEEE.

### Harvard
Jafarzadeh, M., Hussey, D. and Tadesse, Y. (2019). Deep learning approach to control of prosthetic hands with electromyography signals. In: *2019 IEEE International Symposium on Measurement and Control in Robotics (ISMCR)*. IEEE, pp. A1-4-1-A1-4-11.

### Vancouver
Jafarzadeh M, Hussey D, Tadesse Y. Deep learning approach to control of prosthetic hands with electromyography signals. 2019 IEEE International Symposium on Measurement and Control in Robotics (ISMCR). IEEE; 2019. p. A1-4-1-A1-4-11.

### Chicago
Jafarzadeh, Mohsen, Daniel Curtiss Hussey, and Yonas Tadesse. 2019. "Deep Learning Approach To Control Of Prosthetic Hands With Electromyography Signals". In *2019 IEEE International Symposium On Measurement And Control In Robotics (ISMCR)*, A1-4-1-A1-4-11. IEEE.

### BibTeX
@INPROCEEDINGS{ jafarzadeh_hussey_tadesse_2019, author={M. {Jafarzadeh} and D. C. {Hussey} and Y. {Tadesse}}, booktitle={2019 IEEE International Symposium on Measurement and Control in Robotics (ISMCR)}, title={Deep learning approach to control of prosthetic hands with electromyography signals}, year={2019},  pages={A1-4-1-A1-4-11}, doi={10.1109/ISMCR47492.2019.8955725},}

# Deep learning approach to control of prosthetic hands with electromyography signals

Mohsen Jafarzadeh
Department of Electrical and Computer Engineering
The University of Texas at Dallas
Richardson, TX, USA
Mohsen.Jafarzadeh@utdallas.edu

Daniel Curtiss Hussey
Department of Electrical and Computer Engineering
The University of Texas at Dallas
Richardson, TX, USA
dch150330@utdallas.edu

Yonas Tadesse
Department of Mechanical Engineering
The University of Texas at Dallas
Richardson, TX, USA
Yonas.Tadesse@utdallas.edu

*Abstract*—Natural muscles provide mobility in response to nerve impulses. Electromyography (EMG) measures the electrical activity of muscles in response to a nerve's stimulation. In the past few decades, EMG signals have been used extensively in the identification of user intention to potentially control assistive devices such as smart wheelchairs, exoskeletons, and prosthetic devices. In the design of conventional assistive devices, developers optimize multiple subsystems independently. Feature extraction and feature description are essential subsystems of this approach. Therefore, researchers proposed various hand-crafted features to interpret EMG signals. However, the performance of conventional assistive devices is still unsatisfactory. In this paper, we propose a deep learning approach to control prosthetic hands with raw EMG signals. We use a novel deep convolutional neural network to eschew the feature-engineering step. Removing the feature extraction and feature description is an important step toward the paradigm of end-to-end optimization. Fine-tuning and personalization are additional advantages of our approach. The proposed approach is implemented in Python with TensorFlow deep learning library, and it runs in real-time in general-purpose graphics processing units of NVIDIA Jetson TX2 developer kit. Our results demonstrate the ability of our system to predict fingers position from raw EMG signals. We anticipate our EMG-based control system to be a starting point to design more sophisticated prosthetic hands. For example, a pressure measurement unit can be added to transfer the perception of the environment to the user. Furthermore, our system can be modified for other prosthetic devices.

## I. Introduction

The human hand is a very important part, which enables humans to perform basic daily activities ranging from hand gestures to object manipulation. The loss of a hand is a severe mental and physical trauma for humans. It is estimated that there are approximately 94,000 upper limb amputees in Europe [1] and 41,000 upper limb amputees in the United States [2]. The World Health Organization estimates that there are about 40 million amputees in the world [3]. These numbers expected to grow further, due to an increased life expectancy and a corresponding higher incidence of diabetes and vascular diseases. High-performance prosthetic hands significantly improve the quality of life for upper-limb amputees. Passive prosthetic hands [4] are lightweight, robust, and quiet, but can only perform a limited subset of activities. Body-powered [5,6] prosthetic hands are neither intuitive to operate nor they adequately restore limb function. Therefore, developers have investigated externally-powered prosthetic hands for upper-limb amputees for more than a century [7,8]. Recent advances in prosthetic hands aid upper limb amputees to perform some activities that are difficult or impossible when wearing passive devices, up to certain degrees. Control of these prosthetic hands is a key part of developing these devices. Although researchers have proposed many methods to control these hands, there is a huge gap between expected functionality and current state-of-the-art. Thus, control of prosthetic hands is still an open problem. There are several ways to command a prosthetic hand including but not limited to push-buttons, joystick, keyboard, vision, speech, electroencephalography (EEG), electromyography (EMG), and Electroneurography (ENG). Among these ways, electromyography is the most convenient for amputees. Therefore, in this paper, we only focus on controlling the prosthetic hand with electromyography.

Motor neurons transmit electrical signals that cause muscles to contract. Electromyography (EMG) measures muscle response or electrical activity in response to a nerve's stimulation of the muscle. In other words, an EMG electrode reads electric potential generated in the muscle fibers when they contract. There is a strong correlation between the electric potential in the muscle cell and the nerves cell [9]. There are two types of EMG sensors, surface electromyography (sEMg) and intramuscular EMG (imEMG). The imEMG sensors puncture the skin and directly contact to the muscles. Thus, imEMG sensors provide higher signal to noise ratios. Therefore, imEMG sensors are more accurate, especially for motions that involve multiple degrees of freedom of the human hand [10]. However, imEMG sensors are intrusive, painful, uncomfortable, and difficult to set up. Hence, we do not consider imEMG sensors in this paper. SEMG signals are measured with electrodes on the surface of the skin just above the target muscles. So, the sEMGs is a non-invasive technique. Surface EMG sensors typically use silver electrodes and contact with skins can be either wet (silver-chloride) or dry [11]. Wet sEMG sensors have higher signal to noise ratioa. However, the use of a gel is inconvenient for amputees. Here, we focus on the control of prosthetic hand using dry sEMG. For the rest of the paper, we use the term "EMG" word instead of "dry sEMG".

Surface EMG signals have low signal to noise ratios. Therefore, these signals can be easily lost in the environmental noise. The maximum electric potential signal can be measured from sEMG sensors are typically 5 mV [12]. Therefore, in practice, these signals should be amplified (usually 500 to 2000 times) to be able to read with the market available analog to digital converters (ADC). Most EMG sensors in the market use bandpass filter range from 50 to 500 Hz. So, the ADC should be able to measure with the rate of 2000 or more samples per second. The 50 or 60 Hz is easily corrupted with noise because of the power-line in the cities [13]. Some researchers used band-stop filters for the EMG sensors to cancel the 50/60 Hz noise. However, the 50 and 60 Hz bands contain very informative data about the EMG signal. Thus, a functional prosthetic hand should have a low-pass filter (10 Hz) in the power supply circuit to address the power-line noise. In short, designing a measurement circuit with low pass filters (10 Hz) for power-line, amplifiers (500 to 2000 times, typically 660 times), and a/few fast precision ADC (at least 2000 sample per second and preferably 16-bit noise-free resolution) is a key challenge in designing functional prosthetic hands with dry surface EMG sensors. An example of a circuit with 16-bit, 4000 samples per second, third-order Sigma-Delta ADC is demonstrated in [14]. Indeed, the EMG acquisition circuit must be addressed before designing any controller.

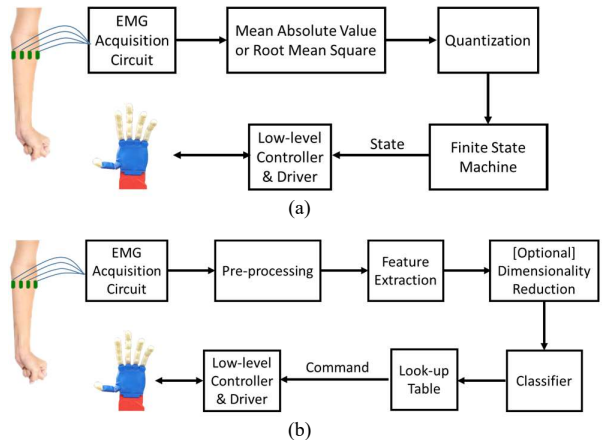

Fig. 1. Traditional approaches to control prosthetics hands with EMG signals (a) using finite-state machine and (b) using signal processing and pattern recognition methods.

There are two traditional ways of controlling devices with EMG signals [15] as shown in Fig. 1. The first one is the quantization of mean absolute value (or root mean square) of EMG signals and using a finite state machine to determine the state of EMG and then use this state to control the device. Although this method is the easiest and most robust way to control the device with EMG, it limits the prosthetic hand to a few basic motions. The second one is using pattern recognition algorithms to classify EMG signals and then using a look-up table to convert the estimated class to a command. The second method typically consists of preprocessing, feature extraction, dimension reduction (feature description), classifier, and a look-up table. The pre-processing may include one or a combination of a high pass filter (10 to 50 Hz), a band-stop filter (50 or 60 Hz), a low pass filter (500 Hz), regression subtraction, spectrum interpolation, non-negative matrix factorization, etc. Feature extraction is the most important subsystem of this system. The feature should be robust to noise while it must discriminate the signals into different classes. A feature extraction system receive a processed EMG signal and computes a feature vector. Researchers have proposed many features [1,16] including but not limited to mean absolute value, integrated absolute value, variance, mean absolute value slop, root mean square, Willison amplitude, number of zero crossing, number of slope sign change, auto-regression coefficients, cepstral coefficient, mean of signal frequency, power spectral density, waveform length, min/max frequency ratio, short-time furrier transform, discrete wavelet transform, and wavelet packet transform. EMG pattern recognition systems typically use 8 or more EMG sensors and 5 or more features (for each sensor).

Therefore, the feature vector has a high dimension. Training of a system with a high dimension features requires a huge amount of data for training, validation, and testing. Therefore, researchers usually use feature description subsystem (dimension reduction techniques) such as principal component analysis (PCA), locality preserving projections (LPP), and neighborhood preserving embedding (NPE). Many EMG classifiers have been studied [1,15] such as support vector machine (SVM), minimax probability machine (MPM), fuzzy neighborhood discriminant analysis (FNDA), k-nearest neighbors (KNN), multilayer perceptron (MLP), self-organized map (SOM), radial basis function (RBF), regulatory feedback networks (RFN), hidden Markov model (HMM), and linear discriminant analysis (LDA). While traditional EMG controllers that use advanced signal processing and pattern recognition appear promising for the recovery of some function in prosthetic hands, still neural control of these devices remains incomplete and unreliable.

Developers have used traditional EMG control systems in many prosthetic and assistive devices. Jang et al. used the EMG signal to control an electric-powered wheelchair [17]. They used root mean square of error and project it to get the state. Gui et al. used RBF classifier for EMG signals to control an exoskeleton [18]. Lyons et al. used linear discriminant analysis to control an upper limb prosthesis by EMG signals[19]. They used a feature vector that consists of mean absolute value, wavelength, number of zero crossings, and number of slope sign changes. Segil et al. used the root mean square of EMG signal with a vector summation algorithm to control the radial motion of a prosthetic hand [20].

Recent advancements in general-purpose graphics processing units (GPGPUs) [21-23] enable devices to run a large set of computations in parallel (real-time). Deep neural networks, especially convolutional neural networks (CNN), benefit from the resulting computation power. Hence, the state-of-the-art optimization methods quickly change from subsystem optimization to end-to-end optimization [24-26]. Image segmentation [27], object detection [28], speech recognition [29], natural language processing [30], playing Atari game [31], object manipulation [32], autonomous driving [33], lung disease detection [34], and financial assert trading [35] are some of the examples of the optimization paradigm shift.

Few studies have been published on application of deep learning for pattern recognition of EMG signals. Atzori et al. archived 70% accuracy by using time windows of 150 ms, preprocessing, and a convolutional neural networks [36]. Their datasets were obtained using 12 electrodes from a Delsys Trigno Wireless System, providing the raw sEMG signal at 2 kHz. These electrodes were fixed on the forearm using standard adhesive bands and a hypoallergenic elastic latex-free band. Eight electrodes were positioned around the forearm at the height of the radio humeral joint at a constant distance from each other; two electrodes were placed on the main activity spots of the flexor digitorum superficialis and of the extensor digitorum superficialis, two electrodes were also placed on the main activity spots of the biceps brachii and of the triceps brachii. Zhai et al. used a CNN consisting of a convolutional layer of and two dense layers [37]. Their CNN input is a spectrogram of EMG signal (i.e. 2D) with window time of 200 ms. Their CNN achieved 78% accuracy. Cote-Allard et al. used a low cost, 8-channel, dry-electrode, low-sampling rate (200 Hz) armband sensor [38]. The data in their work, is separated by applying sliding windows of 260ms (to ensure sufficient time for data processing and classification) with an overlap of 235ms. For each of the eight channels of the armband, 52 samples (260ms) per spectrograms are leveraged using Hann windows of 28 points per fast Fourier transform (FFT) and an overlap of 20. This yields a matrix of 8x4x15 (Channels x Time bins x Frequency bins). Their proposed CNN has three convolutional layers and two dense layers. The few published works on deep learning of EMG signal are not complete. Further study is required for different sensor arrangements and CNN architecture. Also, new methods should be studied on CNN with raw EMG input (instead of spectrogram).

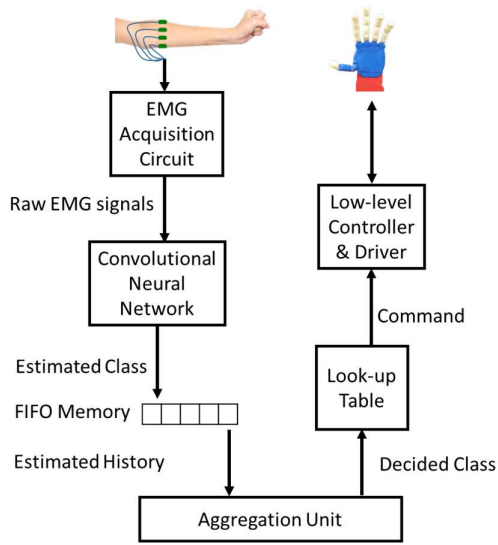

Fig. 2. Block diagram of the proposed method to control prosthetics hands with raw EMG signals using a convolutional neural network.

The main contribution of this paper is the development of a novel EMG based control system for prosthetics hand. The proposed system uses an array of 8 dry, linearly and evenly spaced EMG electrodes. We investigated many CNNs to find the optimal architecture for our sensor arrangement. The proposed convolutional neural networks directly uses raw EMG signals without any preprocessing, feature extraction, or spectrogram. The proposed method is a step toward a fully end-to-end optimization of EMG controlled prosthetic hands. Because the accuracy of the proposed CNN is not 100%, we propose a post-processing subsystem that includes a first input first output (FIFO) memory and an aggregation unit. The proposed post-processing subsystem will make our system error-free while adding a small delay.

This paper is organized as follows. Section II explains the proposed method for EMG-based control of prosthetic hand using deep convolutional neural networks. Section III presents the experimental results of the proposed method when implemented on an embedded GPGPU for prosthetic hands. Section IV explains a preliminary design of electrical circuits. Section V discusses the final products. The last section is a conclusion on the proposed EMG control of prosthetic hands.

## II. METHOD

In this section, we describe a novel method to control prosthetic hands with EMG signals in detail. The main difference between the proposed method and traditional EMG control systems is that instead of four subsystems of pre-processing, feature extraction, dimension reduction (feature description), and a classifier, we use only a novel convolutional neural network. Our novel convolutional neural network uses raw EMG data as input. Fig. 2 shows the block diagram of the system that consists of a novel convolutional neural network, a FIFO memory, an aggregation unit, and a look-up table.

In a study, researchers found that with proper signal processing techniques and pattern recognition systems, the accuracy from signals that captured from non-targeted muscles is close to the accuracy when sensors are exactly positioned above the targeted muscle [39]. Therefore, we decided to use data obtained from an array of eight linearly spaced sensors without concerning about their positioning relative to the muscle groups. While each sensor of this setup captures signals from multiple muscles simultaneously, it is a lot more convenient and simple to set up in repeated uses for amputees.

We decided to develop a novel end-to-end network for estimating the hand gesture class directly from raw EMG signals. The proposed convolutional neural network is novel because, we do not use any pre-processing subsystem and spectrograms. Here, we used 1D convolutional layers instead of 2D convolutional layers. The other novelty of the proposed CNN is that the filter sizes consistently decrease in deeper layers (not fixed). In this paper, we have access to a limited number of datasets such that we are not able to train very deep neural networks (more than 10 layers). As shown in Fig. 3, the proposed networks consist of 6 convolutional layers and 2 dense layers. The input of the networks is 8 vectors, with each vector consisting of 200 elements. The number 200 is the number of sampled data points when the sliding window is 0.1 s and the sampling rate is 2000 samples per second. In this method, we used one-hot encoding. Thus, the output of the network is a vector with number of element equal to the number of class (for example 15 for data set that we used) and the activation function of the last layer is SoftMax. The detail of the networks is stated in Table I.

We proposed to add a post-processing subsystem to overcome the remaining error caused by convolutional neural networks. This subsystem includes first input first output (FIFO) memory that holds the last estimated classes of the CNN and an aggregation unit. The size of FIFO is independent of the hardware (prosthetic hand) that were used. The size of FIFO depends only on the accuracy of the CNN. If someone finds a better CNN that does not have any error, the FIFO memory should be deleted, i.e. size of FIFO will be 1. Size of FIFO should be an odd number. The speed of prosthetic hand decreased linearly by increasing the size of the FIFO memory. Therefore, based on the trained CNN, we should find the minimum odd number such that for all hand gesture in the dataset, the aggregation of all elements in FIFO memory be constant over time when the hand gesture is constant. The CNN output from a stream of EMG signals is independent and identically distributed (iid). Let's denote the probability of misclassification of the CNN with $\rho$ and the size of FIFO memory with $n$. The probability of misclassification after post-processing can be find with $\rho^{\frac{n+1}{2}}$. For example, if the probability of CNN misclassification is 0.1, the probabilities of misclassification after post-processing for FIFO memories with size of 1, 3, and 5 are 0.1, 0.01, 0.001, respectively. Here, we used a FIFO memory with size of 5. As shown in in Fig. 4, the aggregation unit outputs the class that repeated more than half of the size of FIFO. For example for a FIFO with size of 5, output of aggregation unit is a class that repeated 3 times or more. If there is not a class with repetitions more than half of the size of the FIFO memory (here, 3), the output of aggregation unit is the latest estimated class. We used a look-up table to map the decided class (output of the aggregation unit) to a command for the low-level controller of prosthetic hand.

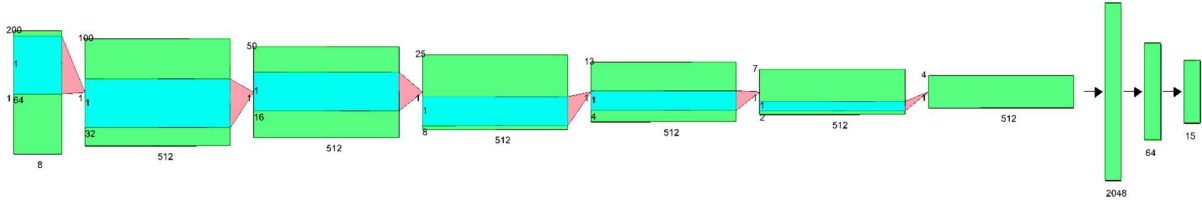

Fig. 3. Block diagram of the proposed convolutional neural network.

TABLE I. DATAILS OF THE PROPOSED CONVOLUTIONAL NEURAL NETWORK.

| Layer | Type | Number of filters | Filter size | Stride | Activation | Output shape | Number of Parameters |
|---|---|---|---|---|---|---|---|
| 0 | Input (Raw EMG) | - | - | - | - | 200 x 8 | 0 |
| 1 | Convolution 1D | 512 | 64 | 2 | ReLU | 100 x 512 | 262656 |
| 2 | Convolution 1D | 512 | 32 | 2 | ReLU | 50 x 512 | 8389120 |
| 3 | Convolution 1D | 512 | 16 | 2 | ReLU | 25 x 512 | 4194816 |
| 4 | Convolution 1D | 512 | 8 | 2 | ReLU | 13 x 512 | 2097664 |
| 5 | Convolution 1D | 512 | 4 | 2 | ReLU | 7 x 512 | 1049088 |
| 6 | Convolution 1D | 512 | 2 | 2 | ReLU | 4 x 512 | 524800 |
| - | Flatten | - | - | - | - | 2048 | 0 |
| - | Drop out | - | - | - | - | 2048 | 0 |
| 7 | Dense | - | - | - | ReLU | 64 | 131136 |
| - | Drop out | - | - | - | - | 64 | 0 |
| 8 | Dense | - | - |  | SoftMax | 15 | 975 |

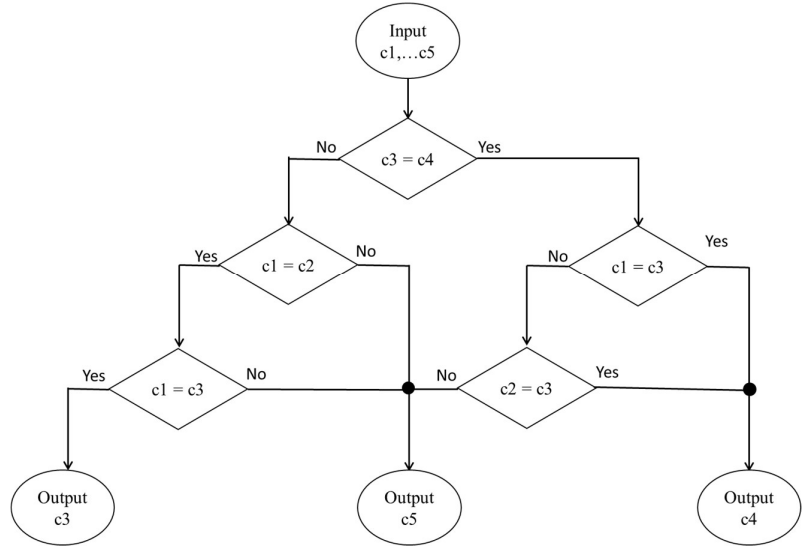

Fig. 4. Flowchart of the proposed aggregation unit for a FIFO memory with size of 5. The c1, …, c5 are predicted classes corresponding to latest CNN outputs.

## III. RESULTS

Selecting EMG dataset is the first steps of doing experiments. We used a dataset consisting 8 subjects, with 3 repetition (20 s long each) of 15 hand positions (total 45 data for each subject). The detail of dataset is stated in the Appendix. We chose this dataset because it uses the proposed sensor arrangements (8 dry, linearly and evenly spaced) and data collected at a higher frequency (4000 sample per second). Then, We down-sampled the data to 2000 sample per second by jumping (not preprocessing). Selecting a high-performance GPGPU for training the networks is the second steps. We chose NVIDIA Tesla V100 GPGPU for our training. Therefore, we used V100 compute nodes in the Maverick2 at the Texas Advanced Computing Center (TACC), The University of Texas at Austin, Austin, TX, USA. A summary of the nodes specification are stated in Table II. The third step to experiment the proposed method is selecting an appropriate developer kit. There are few GPGPU developer kits available in the market that provide GPGPIO, UART, I2C, and SPI buses which are required to communicate with the devices that drive the prosthetic hands. Table III shows a comparison of the two most powerful GPGPU developer kits in the market.

TABLE II. MAVERICK2 V100 COMPUTE NODE SPECIFICATIONS

| Rack | Dell PowerEdge R740 |
|---|---|
| Processors | 2 x Xeon(R) Platinum 8160 CPU 2.10GHz |
| GPGPUs | 2 x NVidia Tesla V100 |
| RAM | 192 GB |
| Cores per processor | 24 |
| Total cores per node | 48 |
| Hardware threads per node | 96 |
| Local storage | 119.5 GB (~32 GB free) |

TABLE III. PROPOSED GPGPU DEVELOPER KITS

| Company | NVIDIA | NVIDIA |
|---|---|---|
| Model | Jetson TX2 | AGX Xavier |
| GPGPU | Pascal 256 core | Volta 512 core |
| TPU | - | - |
| CPU | 4 core Cortex-A57 + 2 core Denver | 8 core Camel |
| RAM | 8 GB | 16 GB |
| Storage | 32 GB | 32 GB |
| GFLOPS | **559** | **1300** |
| GPIO | 8 | 4 |
| USB | 1 x USB 3.0 +1 x USB 2.0 | 2 x USB C [3.1] |
| UART | 1 | 1 |
| I2C | 4 | 2 |
| SPI | 1 with 2 CS | 1 with 2 CS |
| CAN | 1 | 1 |
| I2S | 2 | 1 |
| Size (mm) | 170 x 170 x 51 | 105 x 105 x 85 |
| Weight | 1.5 Kg | 630g |
| Price ($) | 400 | 700 |

NVIDIA AGX Xavier developer kit is the best kit in the market (2019). However, we selected the NVIDIA Jetson TX2 because it provides enough computation power at a lower cost. These two boards are compatible such that users can replace NVIDIA Jetson TX2 with NVIDIA AGX Xavier easily.

The proposed convolutional neural network was implemented in Python 3.5 using the TensorFlow library. We used data from two subjects for testing and other subjects for training and validation. Out of the 3 repetitions of each movement, 2 were chosen and placed into the training set, the other repetition was placed in the validation set. Then all the 3 repetitions from the other 2 individuals were placed into the test set. The input of the proposed CNN is EMG measurement for 100 ms time window. The window slides every 10 ms. Each hand gesture collection was recorderd for 20 s. Thus, the total number of time window in each record is 1991 ($1 + \frac{20-0.1}{0.01}$). Therefore, the whole data consists of 358380 (6 x 15 x 2 x 1991) windows for training, 179190 windows (6 x 15 x 1 x 1991) for validation and, 179190 (2 x 15 x 3 x 1991) windows for testing. We have tried many different architectures of CNN. In our experiments, the eight-layer network (six convolutional layers and two dense layers) shows a validation accuracy of 91.26% (for training accuracy 99.98%). The test accuracy reaches to 48.40%. The training accuracy can be increased by using deeper networks. However, deeper networks result in lower validation accuracies. This is because we used a limited amount of data. We can conclude that by collecting more data and using a deeper network we can achieve better networks.

The results of the training, validation, and test accuracy of the proposed networks is demonstrated in Fig. 5. This CNN has 16,650,255 trainable parameters. The training of this CNN takes 10 hours (200 epochs) in the V100 compute node of Maverick2 supercomputer when the batch size is 4096 and optimizer is Adam (learning rate 0.0001, $\beta_1$=0.9, $\beta_2$=0.999, epsilon=1e-08). The test accuracy is low(48.40%). The main reason for the low test accuracy is the low amount of training data. Nevertheless, the proposed method can be used by fine-tuning of the networks for each new user. Table IV shows a comparison between the proposed CNN and others.

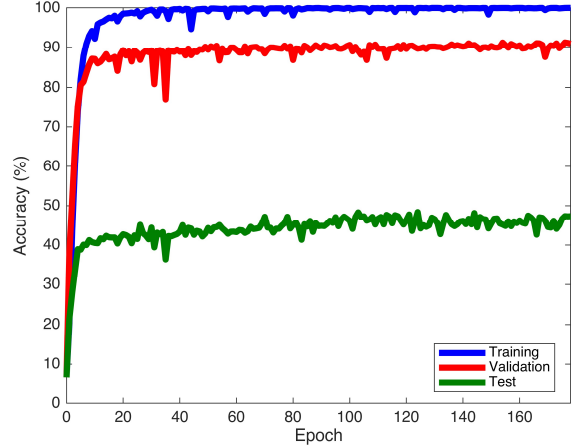

(a)

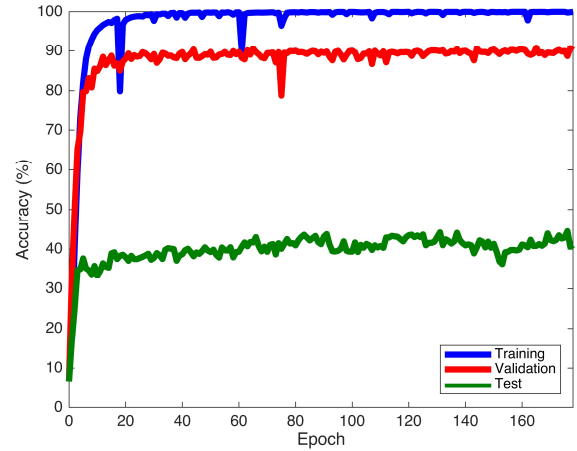

(b)

Fig. 5. Trend of accuracy of the proposed convolutional neural network for two set of cross validation (cross-testing).

TABLE IV. A COMPARISON BETWEEN THE PROPOSED CONVOLUTIONAL NEURAL NETWORKS AND OTHERS

| Method | # sensors | Length (ms) | Input | Accuracy |
|---|---|---|---|---|
| Atzori et al. [36] | 12 | 150 | Preprocessed | 70% |
| Zhai et al. [37] | 12 | 200 | Spectrogram | 78% |
| Cote-Allard et al. [38] | **8** | 260 | Spectrogram | 97.81 |
| Proposed CNN | **8** | **100** | **Raw** | 91.26% |

After training/validating/testing the proposed convolutional neural networks on the supercomputer, we transfer the trained CNN to the embedded GPGPU developer kit (NVIDIA Jetson TX2) for testing. The proposed convolutional neural networks run in real-time. The window slides in 100 ms to match with the desired speed for prosthetic hand low-level controller (10 Hz closed-loop control). We observed that signals are not misclassified in more than two connected windows in each stream by only the CNN, i.e.

used with 14-bit accuracy. This resolution is enough for any prosthetic hands. The voltage regulator circuit consists of an adjustable low dropout linear voltage regulator and a double LC filter.

We used through hole components for the first prototype. However, in the future, we will use surface mount components to reduce noise and overall size. Fig. 7 shows schematics of instrumentation amplifier for the EMG signal, buffer for flex sensor (sensitive resistive), and power amplifier for actuator. Summary of the proposed components is stated in Table V.

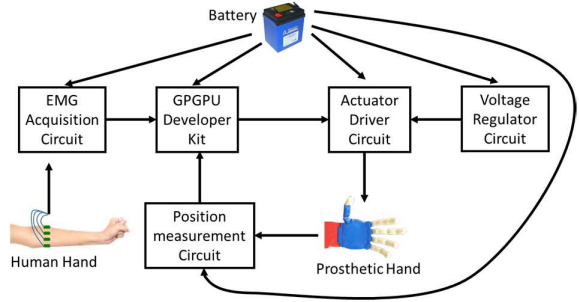

Fig. 6. Hardware diagram of the proposed method for EMG control of a prosthetic hand.

without the post-processing (FIFO memory and aggregation unit). In other words, the joint probability of three consecutive incorrectly estimated classes (CNN without the post-processing) is zero. Therefore, the proposed post-processing makes the final product error-free (functional).

IV. HARDWARE

The hardware of the proposed method consists five subsystems (Fig. 6): EMG acquisition circuit, GPGPU developer kit, actuator driver circuit, position measurement circuit, and voltage regulator circuit. The GPGPU developer kit can be either NVIDIA Jetson TX2 (cost efficient) or NVIDIA AGX Xavier (size efficient and faster). Both EMG acquisition circuit and position measurement circuit have eight 16-bit precision Sigma-Delta analog to digital converters (ADC) with a sampling rate of 2000 samples/second and an ultra-low noise and a low dropout 3.3V linear voltage regulator. To reduce the power input noise, we used a low-pass double LC filter in the both boards. The EMG acquisition circuit uses 8 instrumentation amplifiers (In Amp) with a gain of 660. The position measurement circuit uses 8 single-supply (output rail-to-rail) operation amplifiers in the buffer mode (gain 1) to read flex sensors (bending sensitive resistors). These two boards and the actuator board communicate with GPGPU developer kit via standard I2C bus. We decided to use a separate voltage regulator board for the actuators to have more modularity, thus allowing it to be adapted to a wide range of actuators. The actuator board consists of an 8-channel 16-bit digital to analog converter (DAC), 8 single-supply (output rail-to-rail) operation amplifiers (Op Amp), and 8 NPN bipolar junction transistors (BJT). The actuator board in tandem with regulator boards can provide up to 16V and 2.5 A in each channel, for up to 8 channel). The input power of each channel is independent and can be adjusted for different types of actuators. When the maximum voltage of the actuator is less than 4V, the voltage can be adjusted in the software (hardware adjusted to 4V). In this case the resolution of driver accuracy will drop linearly. For example, 1V actuator can be

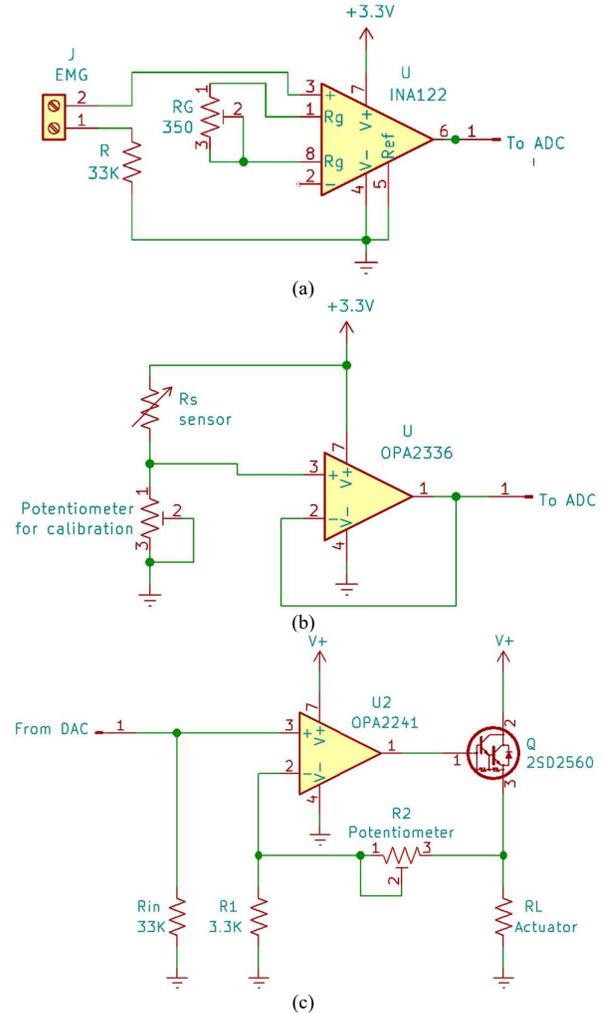

Fig. 7. Schematics of circuit (a) instrumentation amplifier for EMG signal (b) buffer for resistive sensor (such as flex sensor) (c) power amplifier for actuator.

TABLE V. SUMMARY OF PROPOSED COMPOMNENTS FOR FIRST PROTPTYPE

| Components | Model | Description |
| --- | --- | --- |
| DAC | LTC2605IGN-1#PBF | 16-bit, I2C, 8 channels |
| ADC | TI ADS112C04IPWR | 16-bit, I2C, Sigma-Delta, 2000 sample per second |
| I2C Isolator | TI ISO1540QDQ1 | 1Mbps 25kV/μs |

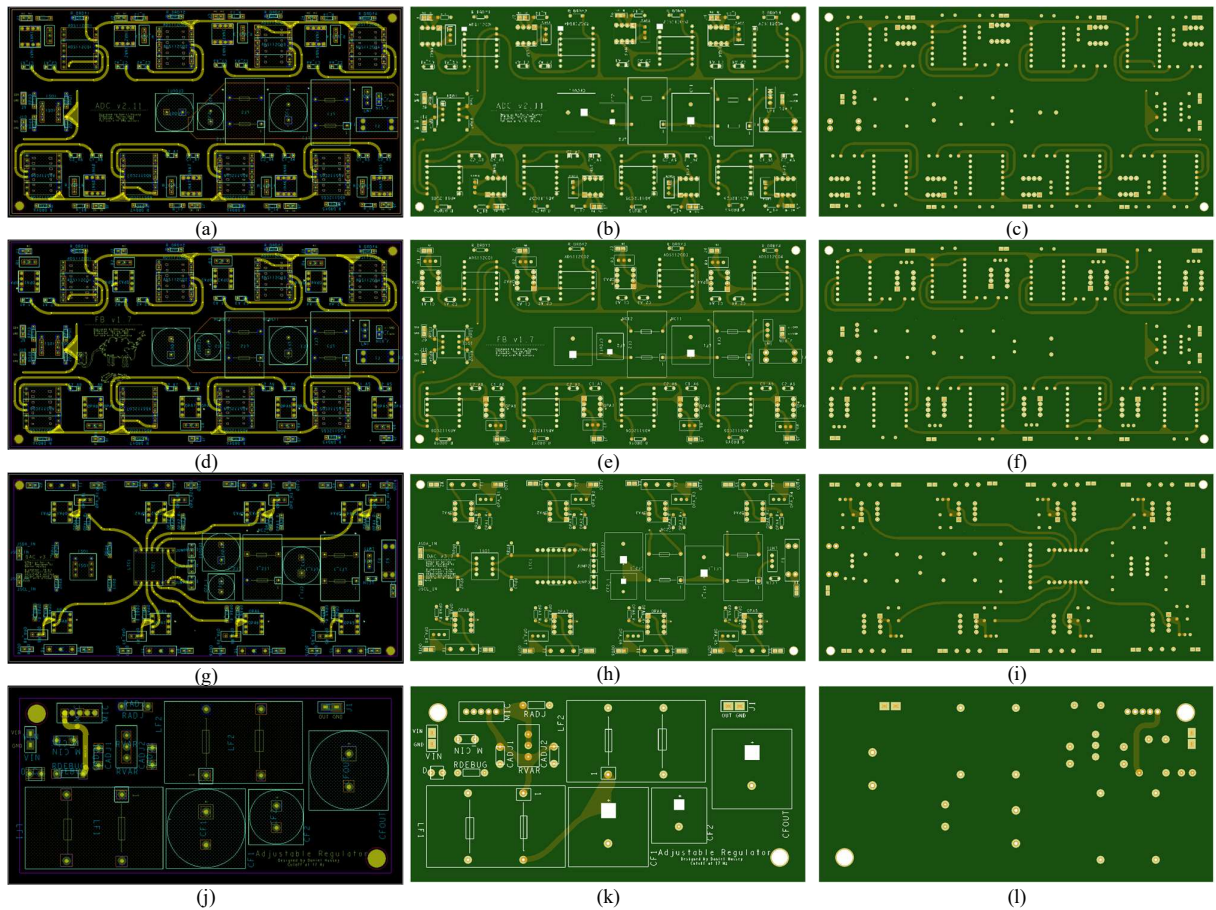

Fig.8. First prototype (a) EMG acquisition circuit CAD model (b) EMG acquisition circuit PCB front (c) EMG acquisition circuit PCB backside (d) position measurement circuit CAD model (e) position measurement circuit PCB front (f) position measurement circuit PCB backside (g) actuator driver circuit CAD model (h) actuator driver circuit PCB front (i) actuator driver circuit PCB backside (j) voltage regulator circuit CAD model (k) voltage regulator circuit PCB front (l) voltage regulator circuit PCB backside.

| In Amp | TI INA122 | Single-Supply, rail-to-rail output, low noise 60nV/√Hz |
|---|---|---|
| Op Amp | TI OPA 2336 | Single-Supply (2.3V to 5.5V), rail-to-rail output (2 mV) |
| Op Amp | TI OPA2241 | Single-Supply (2.7V to 36V), rail-to-rail output (50 mV) |
| BJT | Sanken 2SD2560 | NPN, DC Current Gain 5000, Power Max 130W |
| Adjustable linear voltage regulator | LT1764AET#06PBF | Output 3A, dropout 0.34V, low noise 40μVRMS |
| 3.3 V linear voltage regulator | TI LM1085IT-3.3/NOPB | load regulation 0.1%, 3A, low noise 38μVRMS |

The CAD model and PCB of the first prototypes are illustrated in Fig. 8. The PCBs have been designed to be fabricated in green color, 2.0 mm thickness, 2 oz copper weight, and ENIG-RoHS surface finish.

## V. DISCUSSIONS

In this paper, we used EMG data from 6 people for training/validation and two people for testing of the proposed network. The small gap between the training accuracy and the validation accuracy (Fig.5) shows that the proposed method is reliable (functional) when this CNN is fine-tuned for a specific person. However, there is a huge gap between the validation accuracy and the test accuracy. This large magnitude shows that the proposed method cannot be generalized to work for many different individuals. The first step in the future is increasing the number of people in the dataset from 8 to 100 or more in order to study the robustness (test accuracy). Another step will be investigating very deep convolutional neural network, which requires increasing the number of people to 1000 or even higher.

Here, we proposed the aggregation unit to overcome the small amount of the error in validation. The alternative way to the proposed post-processing subsystem is using a recurrent layer such as long short-term memory (LSTM) and gated recurrent unit (GRU). The recurrent neural networks consider feedback from past decisions. However, the resulting networks require a higher amount of data. Thus, we will study these types of networks after collecting more data.

In this study, we used only the supervised classification learning method for application in prosthetic hands. The main disadvantage of the proposed method is that it is limited to a set of predefined positions (here 15 gestures). Imitation learning and reinforcement learning are other types of the learnings that can be used for continuous control of the hand (not a discrete finite set of predefined gesture). In the future, after collecting more data, we will study imitation learning and reinforcement learning technique for continuous control of prosthetic hands.

In this research, we focused on a method of controlling prosthetic hands using raw EMG signal as an input with deep convolutional neural networks. However, our methods can be implemented on the other prosthetic devices. The only difference is optimizing the architecture (layers) of convolutional neural networks for those particular devices. The only change will be update the mapping subsystem (look-up table).

## VI. CONCLUSIONS

We used a dataset that was obtained from 8 people with 15 hand gestures and 3 repetitions. Two of the three repetitions from 6 people are used for training, the third repetitions of 6 people are used for validation. All the signals from the other two persons were used for testing. The validation accuracy of the proposed method is 91.26%. Because the accuracy of the proposed convolutional neural network is not 100%, we propose a post-processing subsystem that includes a first input first output (FIFO) memory and an aggregation unit. The proposed post-processing subsystem makes our system error-free.

In the future, first, we will collect more EMG data from many people. After that, we will be able to study deeper neural networks. The low-test accuracy (48.40%) can be addressed by a higher amount of data. Then, we will compare recurrent neural networks with the proposed post-processing subsystem. The main disadvantage of the proposed method is that it is limited to a finite number of hand gestures (here 15). In the future, we will study deep imitation learning and deep reinforcement learning to address continues control of prosthetic hands with EMG signals.

## APPENDIX: DATASET

Data is the most important part of any machine learning algorithm, especially deep learning. The proposed method uses an array of 8 EMG electrodes and raw EMG data for the neural network. We used the data set that was published by Khushaba and Kodagoda [40]. Here, we will explain how the data was recorded and the nature of the experiment as this data

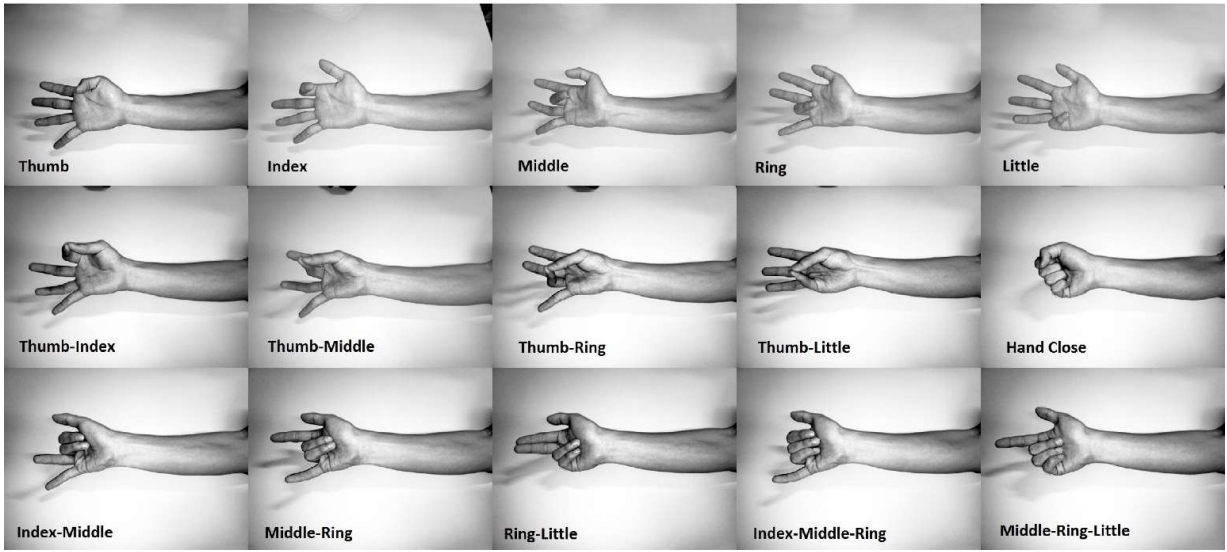

Fig. 10. Different classes of individual and combined fingers movement [40].

In this paper, we use deep learning techniques to develop a novel EMG based control system for prosthetics hands with an array of 8 dry, linearly and evenly spaced surface EMG electrodes (2000 sample per second). In contrast to recent literature, the proposed convolutional neural network uses raw EMG (with window time of 0.1s) without any preprocessing (including a spectrogram). The proposed method was implemented in Python using the TensorFlow library. We use an NVIDIA Jetson TX2 developer kit (an embedded GPGPU board) to test the proposed methods.

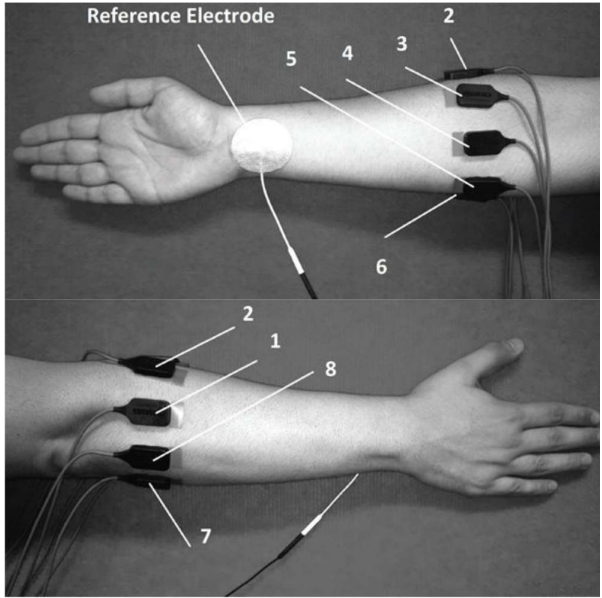

Fig. 9. EMG sensor arrangement [40].

is input to our network. In their project, eight subjects, six males and two females, aged between 20-35 years were recruited to perform the required fingers movements. The subjects were all able body subjects with no neurological or muscular disorders. Subjects were seated on an armchair, with their arm supported and fixed at one position. The datasets were recorded using eight EMG channels (DE 2.x series EMG sensors) mounted across the circumference of the forearm and processed by the Bagnoli desktop EMG system from Delsys Inc., as shown in Fig. 9. Fifteen classes of movements were collected including: the flexion of each of the individual fingers, i.e., Thumb, Index, Middle, Ring, Little and the combined Thumb-Index, Thumb-Middle, Thumb-Ring, Thumb-Little, Index-Middle, Middle-Ring, Ring-Little, Index-Middle-Ring, Middle-Ring-Little, and finally the hand close class as shown in Fig. 10. Subjects were asked to perform 3 repetitions of the 15 different hand gestures. The data was captured with a rate of 4000 sample per second. We down-sampled to 2000 by jumping data points (not preprocessing). Fig. 11. Illustrates an input to the proposed convolutional neural network, which was a measurement result obtained from [40]. Table VI is the look-up table for this data set. The value (command) of a finger in the prosthetic hands is zero when the finger is fully relaxed. The value is one when the finger is fully contract.

TABLE VI. LOOK-UP TABLE FOR THE DATAST SHOWN IN FIG. 10.

| Hand Gesture | Thumb Value | Index Value | Middle Value | Ring Value | Pinky Value |
|---|---|---|---|---|---|
| Thumb | 1 | 0 | 0 | 0 | 0 |
| Index | 0 | 1 | 0 | 0 | 0 |
| Middle | 0 | 0 | 1 | 0 | 0 |
| Ring | 0 | 0 | 0 | 1 | 0 |
| Little | 0 | 0 | 0 | 0 | 1 |
| Thumb-Index | 1 | 1 | 0 | 0 | 0 |
| Thumb-Middle | 1 | 0 | 1 | 0 | 0 |
| Thumb-Ring | 1 | 0 | 0 | 1 | 0 |
| Thumb-Little | 1 | 0 | 0 | 0 | 1 |
| Hand Close | 0 | 0 | 0 | 0 | 0 |
| Index-Middle | 0 | 1 | 1 | 0 | 0 |
| Middle-Ring | 0 | 0 | 1 | 1 | 0 |
| Ring-Little | 0 | 0 | 0 | 1 | 1 |
| Index-Middle-Ring | 0 | 1 | 1 | 1 | 0 |
| Middle-Ring-Little | 0 | 0 | 1 | 1 | 1 |

ACKNOWLEDGMENT

We would like to thank Cameron Ovandipour and Ngoc Tuyet Nguyen Yount for valuable contributions in developing this project. We would like to express our very great appreciation to Dr. Marco Tacca, Dr. Nicholas Gans, Dr. Neal Skinner, and Dr. John Hanson, for comments and guidance that greatly improved the projects. We would also like to show our gratitude to Dr. Simmons and the Texas Advanced Computing Center (TACC), The University of Texas at Austin, Austin, TX, USA, for providing computational resources, (Maverick2 supercomputer).

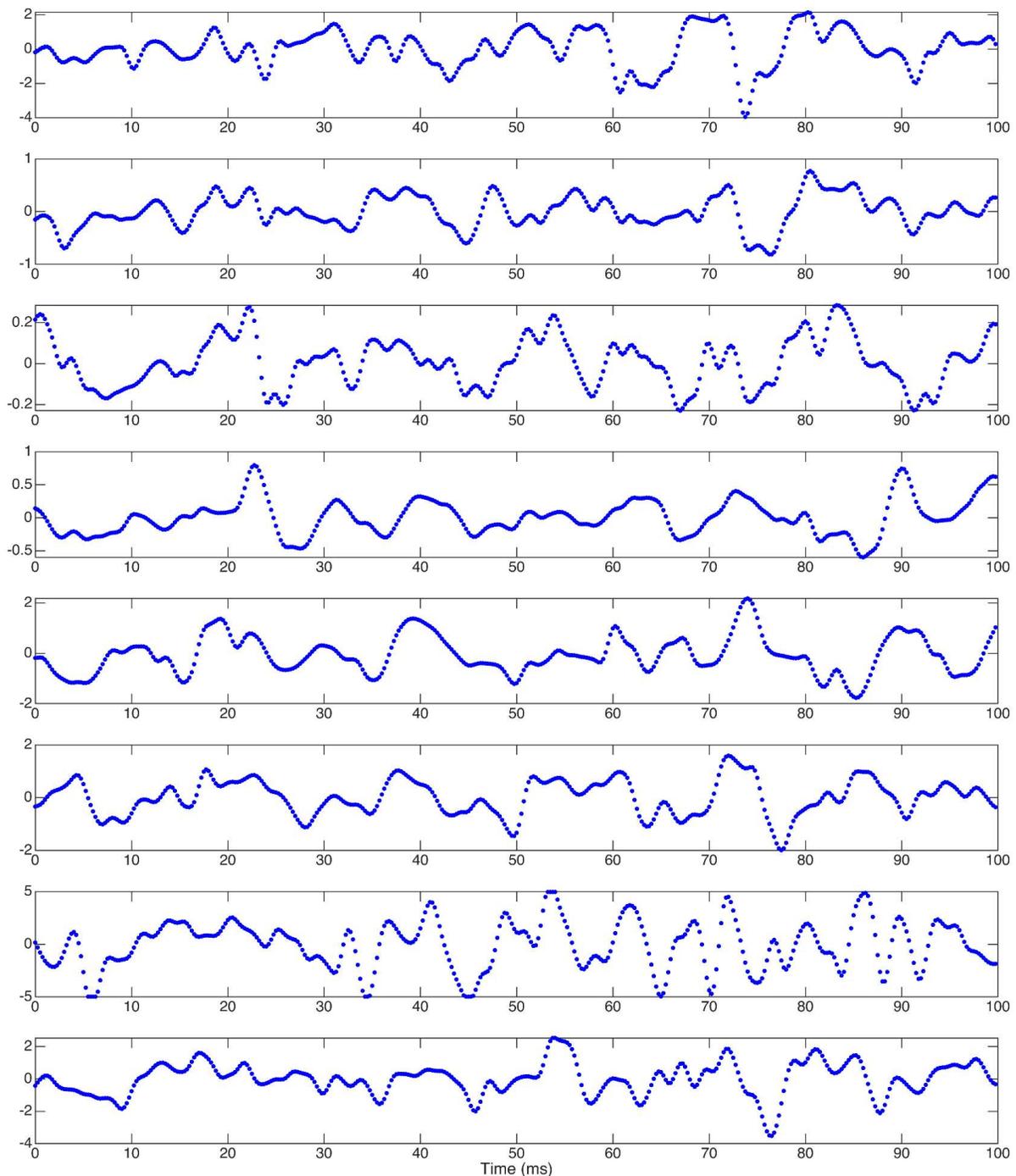

Fig. 11. Raw EMG signals obtained from human subject, which is one example of input data used in our convolutional neural network. The input data is in mV (raw EMG signals, 8 channels).

**DOI:** 10.1109/ISMCR47492.2019.8955725

**Link:** https://ieeexplore.ieee.org/document/8955725

Cite as:

**IEEE**
M. Jafarzadeh, D. Hussey and Y. Tadesse, "Deep learning approach to control of prosthetic hands with electromyography signals", in *2019 IEEE International Symposium on Measurement and Control in Robotics (ISMCR)*, Houston, Texas, USA, 2019, pp. A1-4-1-A1-4-11.

**ACM**
Jafarzadeh, M., Hussey, D. and Tadesse, Y., 2019. Deep learning approach to control of prosthetic hands with electromyography signals. In *2019 IEEE International Symposium on Measurement and Control in Robotics (ISMCR)*. IEEE, pp. A1-4-1-A1-4-11.

**MLA**
Jafarzadeh, Mohsen et al. "Deep Learning Approach To Control Of Prosthetic Hands With Electromyography Signals". IEEE, *2019 IEEE International Symposium On Measurement And Control In Robotics (ISMCR)*. IEEE, 2019, pp. A1-4-1-A1-4-11.

**APA**
Jafarzadeh, M., Hussey, D., & Tadesse, Y. (2019). Deep learning approach to control of prosthetic hands with electromyography signals. In *2019 IEEE International Symposium on Measurement and Control in Robotics (ISMCR)* (pp. A1-4-1-A1-4-11). Houston, Texas, USA: IEEE.

**Harvard**
Jafarzadeh, M., Hussey, D. and Tadesse, Y. (2019). Deep learning approach to control of prosthetic hands with electromyography signals. In: *2019 IEEE International Symposium on Measurement and Control in Robotics (ISMCR)*. IEEE, pp. A1-4-1-A1-4-11.

**Vancouver**
Jafarzadeh M, Hussey D, Tadesse Y. Deep learning approach to control of prosthetic hands with electromyography signals. 2019 IEEE International Symposium on Measurement and Control in Robotics (ISMCR). IEEE; 2019. p. A1-4-1-A1-4-11.

**Chicago**
Jafarzadeh, Mohsen, Daniel Curtiss Hussey, and Yonas Tadesse. 2019. "Deep Learning Approach To Control Of Prosthetic Hands With Electromyography Signals". In *2019 IEEE International Symposium On Measurement And Control In Robotics (ISMCR)*, A1-4-1-A1-4-11. IEEE.

**BibTeX**
```
@INPROCEEDINGS{ jafarzadeh_hussey_tadesse_2019, author={M. {Jafarzadeh} and D. C. {Hussey} and Y. {Tadesse}},
booktitle={2019 IEEE International Symposium on Measurement and Control in Robotics (ISMCR)}, title={Deep learning approach to control of prosthetic hands with electromyography signals}, year={2019},  pages={A1-4-1-A1-4-11},
doi={10.1109/ISMCR47492.2019.8955725},}
```